\documentclass[twocolumn,showpacs,preprintnumbers,amsmath,amssymb]{revtex4}
\usepackage{graphicx}% Include figure files
\usepackage{dcolumn}% Align table columns on decimal point
\usepackage{bm}% bold math
\usepackage{bm}

\usepackage{cancel}
\usepackage{epsfig}
\newcommand{\beq}{\begin{equation}}
\newcommand{\eeq}{\vspace{0cm} \end{equation}}
\newcommand{\beqq}{\setlength\arraycolsep{2pt}\begin{eqnarray}}
\newcommand{\eeqq}{\vspace{0cm} \end{eqnarray}}

\begin{document}

\title{Gravitational Matter Creation, Multi-fluid Cosmology and Kinetic Theory}
\author{S. R. G. Trevisani\footnote{strevisanijr@usp.br}}
\author{J. A. S. Lima\footnote{jas.lima@iag.usp.br}}
\affiliation{Departamento de Astronomia, Universidade de S\~{a}o
Paulo \\ Rua do Mat\~ao, 1226 - 05508-900, S\~ao Paulo, SP, Brazil}

%\affiliation{$^{2}$Departamento de F\'isica Geral, Universidade de S\~{a}o
%Paulo \\ Rua do Mat\~ao, 187 - 05508-090, S\~ao Paulo, SP, Brazil}

%\keywords{Dark energy, cosmic distance, Inhomogeneity Parameter}

\bigskip
\begin{abstract}

 A macroscopic and kinetic relativistic description for a decoupled multi-fluid cosmology endowed with  gravitationally induced particle production of all components is proposed. The temperature law for each decoupled particle species is also kinetically  derived. The  present approach points to the possibility of an exact (semi-classical) quantum-gravitational kinetic treatment by incorporating back reaction effects for an arbitrary set of dominant decoupled components.  As an illustration we show that a cosmology driven by creation of cold dark matter and baryons (without dark energy) evolves like $\Lambda$CDM. However, the complete physical emulation is broken when  photon creation is added to the mixture thereby pointing to a crucial test in the future. The present analysis also open up a new window to investigate the Supernova-CMB tension on the values of $H_0$, as well as the $S_8$ tension since creation of all components changes slightly the CMB results and the expansion history both at early and late times. Finally, it is also argued that cross-correlations between CMB temperature maps and the Sunyaev-Zeldovich effect may provide a crucial and accurate test confronting extended CCDM and $\Lambda$CDM models.
\end{abstract}

\maketitle

\section{Introduction}

The late time accelerating stage of the universe is usually explained by assuming the existence of a dominant dark energy (DE) component, in addition to cold dark matter (CDM) and baryons. Its  most popular candidate is the cosmological constant ($\Lambda$) or the rigid energy density of the current false vacuum state ($\rho_V=\Lambda/8\pi G$). The observational pillars providing convincing evidences for the so-called $\Lambda$CDM model include several independent astronomical observations \cite{A2019,Planck2018}. When combined with the primeval inflation for describing the first stages of the early universe including the resulting scenario (inflation + $\Lambda$CDM) is widely known to be considerably simple and quite predictive.
%the formation of the thermal bath, the causal origin of the seeds required by the structure formation process, and also the solution to the problems of horizon and the quasi-flatness of the Universe, the resulting scenario (inflation + $\Lambda$CDM) is not only impressive but also quite predictive and considerably simple \cite{A2019,Planck2018}. 
%Supernovae Ia, Baryon Acoustic Oscillation (BAO), and the processed matter and angular power spectrum of the cosmic background radiation (CMB) temperature anisotropies.  
%In addition, by adopting  the primeval inflation for describing the first stages of the early universe including the formation of the thermal bath, the causal origin of the seeds required by the structure formation process, and also the solution to the problems of horizon and the quasi-flatness of the Universe, the resulting scenario (inflation + $\Lambda$CDM) is not only impressive but also quite predictive and considerably simple \cite{A2019,Planck2018}. 

Nevertheless, there are two old theoretical cosmological puzzles or mysteries plus at least two recent observational difficulties plaguing the $\Lambda$CDM model, namely: (i) the cosmological constant problem \cite{SW89}, (ii) the coincidence problem \cite{C1},  (iii)  the statistical observational discrepancy between measurements of the Hubble constant ($H_0$) from Supernovae (SNe) and other distance indicators at low \cite{LV19} and intermediate redshifts \cite{JV07} as compared  with independent estimates at high redshifts based on the CMB angular power spectrum, and (iv)  the so-called $S_8$ tension on the ($\sigma_8, \Omega_M$) plane by confronting  Planck + $\Lambda$CDM estimates with cosmic shear experiments \cite{Sigma8a},  where $\sigma_8$ measures the current mass fluctuation in a scale of 8$h^{-1}$Mpc. Currently (both tensions $H_0$ and $S_8$  are the major observational anomalies plaguing the $\Lambda$CDM model (see below).
%a tension of the Planck data with weak lensing measurements and redshift surveys has
%been reported, about the value of the matter energy density 

Many attempts to solve or alliviate the theoretical puzzles gave rise to a plethora of dark energy possibilities including different kinds of running vacuum or decaying $\Lambda$-models, interactions in the dark sector and other noncanonical scalar fields \cite{L1992,WZ00,Lima04,AL2005,Sahni06,Sam2006,PJ09,SD2015,ZSL2018,LL2018,LA2021}. There are  also more fundamental approaches beyond Einstein's theory, like several  extensions of Einstein's general relativity, among them: F(R), F(R,T) and  Gauss-Bonnet type theories \cite{L1,L2,Harko2,C2011,GL20}. 

In the observational front, Riess and collaborators are now claiming for a statistical discrepancy of $5\sigma$ level between the local $H_0$ value and the one  predicted by Planck + $\Lambda$CDM \cite{Riess21}. 
%They also argued against the presence of possible  non-negligible  measurement errors and a large set of analysis variations considered to date. 
This means that CMB-SNe tension remains unsolved regardless of the realistic dark energy model in general relativity.  Further, although  statistically less significant ($2.6\sigma$ to 3$\sigma$ confidence levels) in comparison with the $H_0$ trouble,  the $S_8$ estimates based on cosmic shear measurements from weak lensing collaborations, like Kilo-Degree Surveys (KiDS) and the Dark Energy Survey (DES) are providing values for the parameter $S_8 = \sigma_8\sqrt{\Omega_M/0.3}$  lower than the early-time probes \cite{KiDS20,DES18,M2021}. Such observations are clearly opening the possibility to cosmic scenarios beyond $\Lambda$CDM model. Actually, some authors are claiming that solutions for the $H_0$ and $S_8$ tensions may require changes in the expansion history both at early and late-times (for more details see \cite{EV2021a,EV2021b}).

In this context, the central interest here is to investigate a possible reduction of the dark sector by eliminating within general relativity, all possible species of dark energy, that is, we set $\Omega_{DE}\equiv 0$ from the very beginning, including the rigid vacuum itself. Therefore, even if early inflation was caused by the dominance of a vacuum state, its energy density was totally spent to create light particles forming the primeval thermal bath, as usually assumed in many spontaneously symmetry breaking models \cite{KT90,M2005}. In addition,  any subsequent phase transition was also unable to generate a sizable vacuum state potentially capable to accelerate the universe at late times. This means that the alluded discrepancy of the $\Lambda$-term and also the coincidence problem would be naturally solved. 
%assuming $\Omega_{DE}\equiv 0$. 

In scenarios with $\Omega_{DE}=0$ new challenges take place. For instance, some mechanism emulating the late time accelerating $\Lambda$CDM evolution must to be proposed at the level of the  cosmic smooth expansion. Further, any new picture must also be successful in the perturbative approximation. In other words, although only slightly different from $\Lambda$CDM evolution, it needs to be as close as possible to the perturbed $\Lambda$CDM description. 

The mechanism adopted here is the gravitationally induced particle creation by the expanding  universe, a process already investigated in general relativity  and also in alternative theories of gravity. Such investigations were carried out both from microscopic and macroscopic viewpoints. The former is based on methods and techniques from quantum field theory in curved spacetimes \cite{QFT1,QFT2,PBL2010,CAP16}, while the latter rested upon the non-equilibrium irreversible thermodynamic approach \cite{P89,CLW92,Harko2}. Here  we focus our attention on the irreversible macroscopic and its associated relativistic kinetic counterpart. The basic reasons are briefly outlined below. 

%The difference between the scalar irreversible processes of matter creation and bulk viscosity was also clarified \cite{LG92}.  
Some early theoretical attempts \cite{LRW96,LSS09} gave rise a decade ago to a new accelerating cosmology based on the ``adiabatic" creation of cold dark matter (CCDM) \cite{LJO2010}.  In this general relativistic  model with ($\Omega_{DE}=0$), the cosmic smooth history is fully equivalent to the $\Lambda$CDM model. This very compelling aspect  is not shared by any previous phenomenological matter creation models. In particular, the transition from a decelerating  to the late-time accelerating stage happens at the same redshift. The evolution of perturbations was also discussed in such framework \cite{Pert}.  Under certain circumstances the CCDM  dynamics is equivalence to the $\Lambda$CDM cosmology  not only at the level of the Hubble flow but also for the evolving matter fluctuating field.  Actually, it was demonstrated that the  CCDM cosmology (without creation of baryons) emulates perfectly the $\Lambda$CDM model in the linear and nonlinear levels \cite{Waga2014a,Waga2014b}.  Moreover, a kinetic approach for a single component based on a modified relativistic Boltzmann equation with matter creation was also proposed and the CCDM cosmology was kinetically recovered \cite{LB2014}. Later on, a model with creation of non-relativistic components (baryons + cold dark matter) with different creation rates was also proposed \cite{LSC2016}. This scenario was also proved to be equivalent to $\Lambda$CDM also at a perturbative level, thereby confirming in a more general way the results of Ref. \cite{Waga2014a}.  

CCDM type models have also been tested through a Bayesian analysis applied to SNe Ia data and clusters.  A joint analysis (without creation of photons) involving baryon acoustic oscillations (BAO) + cosmic microwave background (CMB) + SNe Ia data yielded $\Omega_m=0.28 \pm 0.01 (1\sigma)$, where $\Omega_m$ is the matter density parameter. In particular, this implies that the model has no dark energy but the part of the matter that is effectively clustering is in good agreement with determinations from the large-scale structure \cite{JVA2017}. 

It is also interesting that the simplest extensions of the original CCDM model by including baryons, mimicks exactly the observed accelerating $\Lambda$CDM cosmology with just one dynamical constant free parameter $\Gamma = \Gamma_b + \Gamma_{dm}$ describing the total creation rate of both components.  Since the  model is also equivalent to $\Lambda$CDM at the perturbative levels, it reinforces the idea that the  ``cosmic concordance model" may be only an effective cosmology. However, this macroscopic non-equilibrium treatment was not the most general one since the behavior of the CMB radiation with creation was separately discussed \cite{LTS2021}, and, as such, not properly inserted in the complete picture. In principle, the thermodynamic and kinetic results for massless particles remain valid even for dark photons and massless dark fermions \cite{ABCK09}.

Here we explore this kind of model one step further by discussing the general macroscopic formulation for a decoupled multi-fluid mixture endowed with ``adiabatic and\, ``non-adiabatic" matter creation of all components. It will be demonstrated here that the  most interesting kinetic counterpart for applications to late time cosmology is the ``adiabatic" case. Such multi-fluid formulations (macroscopic and kinetics) provide a detailed and coherent extension of the partial results discussed in the above quoted papers thereby suggesting a new route to investigate the $H_0$ and $S_8$ tensions, and, naturally, the CMB anisotropies and distortions. As far as we know, this is the first detailed study proposing the basic complete approach (macroscopic and kinetic) for a decoupled mixture with creation of all components. 
%In a subsequent paper (paper II), the first-order kinetic version of the present approach (the hierarchy of disturbed Boltzmann modified equations) for photons in terms of the corresponding creation rate. The same will be done to the other relevant components (CDM, baryons and neutrinos) including the tightly coupled limit and solutions for the baryon-photon fluid. Such results are indispensable to run the codes for obtaining the complete set (power spectrum) of CMB anisotropies with creation but, naturally, are out of the scope of the present paper. 

The present article is planned as follows. In section {\bf II}, the cosmic irreversible particle production process for a multi-fluid with different creation rates is thermodynamically and dynamically formulated. Corrections for the dynamic pressure and temperature law for each component are  deduced assuming different rates for the particle and entropy productions, but a special attention will be physically justified  for the so-called ``adiabatic" creation process. In section {\bf III}, a Boltzmann equation for mixtures with ``adiabatic" creation is proposed  (some technicalities related to the extended Boltzmann equation are presented in the Appendix A). In section {\bf IV}, the  counterpart of all non-equilibrium macroscopic results  are  kinetically derived. As an example of both consistent routes for a multifluid description (irreversible thermodynamics and kinetics),  subsection {\bf VA} is dedicated to a new extended CCDM cosmology including creation of baryons, CDM, CMB photons and neutrinos with different creation rates, whereas in {\bf VB} and {\bf VC} we focus on  CMB effects (distortions and temperature anisotropies) with emphasis in observations relating distortions and CMB temperature maps. In particular, the cross-correlation  of  SZE  and the integrated Sachs-Wolfe (ISW) effect,  is suggested here as a crucial and accurate test for confronting CCDM and $\Lambda$CDM models. Finally, the article is closed in section VI, by summarising  the main results and conclusions for the relativistic extended accelerating model without dark energy ($\Omega_{DE}\equiv 0$) powered by gravitationally induced particle production.  
%an accelerating scenario without dark energy, but endowed with creation of two components (dark matter, baryons) is briefly presented in section VI. The influence of the CMB component and neutrinos is also discussed. 
%It is also demonstrated that the equilibrium distribution for non-quantum massive and massless particles are preserved for all components in the decoupled mixture.
%Finally, the article is closed in section VI, by summarising the basic conclusions and final comments. Unless explicitly stated, in this work we adopt units with $\hbar = k_B=c=1$.

\section{Irreversible particle production: Macroscopic multi-fluid formulation}

For the sake of simplicity,  let us consider that the spacetime geometry is described by a flat ($k=0$) Friedman-Lemaître-Robertson-Walker (FLRW) metric:

\begin{equation}\label{FRW}
 ds^2 = dt^2 - a^{2}(t)\left[dx^2 + dy^2  + dz^{2}\right],
\end{equation}
where $a(t)$ is the scale factor. For this metric, the non-null Christoffel's symbols are:

\begin{equation}\label{Csymb}
\Gamma^{0}_{00}=\Gamma^{i}_{jk}=0, \,\,\,\Gamma^{i}_{0j}= H\delta^{i}_{j},\,\,\, \Gamma^{0}_{ij} = -Hg_{ij},  \,\,\,
\end{equation}
where $\dot H = {\dot a}/a$ is the Hubble parameter.

The above expanding FLRW  geometry (\ref{FRW}) is also assumed capable to produce all species of particle existing in the Universe. In principle, due to the time-varying gravitational field all the particle components are springing-up in the spacetime with different creation rates. 
%($\Gamma_{iN}$, $i=1,2,....N$).  
As will be discussed next, this macroscopic statement is also in agreement with the standard quantum field theoretic approach with an advantage, namely: the back reaction effect on the geometry is introduced from the very beginning through a specific stress term. 
%In this section, based on the macroscopic formulation for a single fluid (see for instance, Refs. \cite{LJO2010,LTS2021}), we propose a rigorous macroscopic description for a multi-fluid mixture. Our approach also recover as a particular case the  results derived in the above quoted papers. The associated multi-fluid kinetic formulation will be postponed to the next section.

Following standard lines, the non-equilibrium thermodynamic states of a relativistic comoving mixture may be characterised by $3N$-independent macroscopic quantities associated to the distinct components: the energy-momentum tensor (EMT), $T^{\mu \nu}_{(i)}$, a particle current, $N_{(i)}^{\mu}$, and the entropy current, $S_{(i)}^{\mu}$, where the total quantities are summed over all species:

\begin{equation}
T^{\mu\nu}_{T}=\sum_{i=1}^{N}T^{\mu \nu}_{(i)},\,\,\,\,\,\,N^{\mu}_{T}=\sum_{i=1}^{N}{N}^\mu_{(i)},\,\,\,S^{\mu}_{T}=\sum_{i=1}^{N}S^{\mu}_{(i)}. 
\end{equation}
where $i=1,2, ... ,N$ denotes the  $i$-th fluid component in the mixture. 

Now, irreversible particle creation requires a modification of the basic  equilibrium equations. In order to clarify how the above basic thermodynamic fluxes are  modified, we need to modify  the EMT and the balance equations for the particle number density and entropy currents in agreement with the second law of thermodynamics. Let us first consider the possible corrections in the EMT. It can be written as:

\begin{equation}\label{correction1}
T_{(i)}^{\mu\nu} =  {T_{(i)|E}^{\mu\nu}} + \Delta T_{(i)}^{\mu\nu},
\end{equation}
where ${T_{(i)|E}^{\mu\nu}}$ describes the equilibrium states and  $\Delta T_{(i)}^{\mu\nu}$ is the correction associated to the effects of particle production. The homogeneity and isotropy of the FLRW metric implies that the only possibility is a scalar process which in terms of components reads:  
\begin{equation}\label{Comp}
{\Delta T^{0}_{0(i)}} = 0 \,\,\, \text{and} \,\,\, \Delta {T^{i}_{j(i)}}=P_{c(i)} \delta^{i}_{j},
\end{equation}
where Latin indexes in the second equality containing round brackets are not summed. Note also that the first condition (for each component) removes the ambiguity on the energy density for non-equilibrium states. It means that $\rho_i$ is the same function of the thermodynamic variables in the absence of dissipation, and $P_{c(i)}$ is a dynamic pressure here describing macroscopically the emergence of particles into the spacetime. Although similar, it cannot be confused with the scalar (collisional) process of the standard nonequilibrium fluid mechanics and kinetic theory, widely known as bulk viscosity (second viscosity). In a manifestly covariant description we can write

\begin{equation}\label{deltaT}
 \Delta T_{(i)}^{\mu\nu} =  P_{ci} (g^{\mu \nu} - u^\mu u^\nu)\equiv -P_{ci} h^{\mu\nu}, 
\end{equation}
where $h^{\mu\nu}$ is the projector onto the rest frame of $u^\mu$. As happens in the nonequilibrium thermodynamics, the correction for each component, $\Delta T_{(i)}^{\mu\nu}$, works like a source term for the equilibrium EMT. It can be incorporated back for describing the whole process through a conserved EMT as required by the Einstein field equations. Finally, by extending the one-fluid description and assuming for a while that at late times all components filling  the universe are decoupled even in presence of gravitationally induced matter creation, the EMT of each component takes the form: 
\begin{equation}\label{EMT1}
{T_{(i)}^{\mu\nu}} = (\rho_i + p_i + P_{ci})u^{\mu} u^{\nu} - (p_i + P_{ci}) g^{\mu \nu}.
\end{equation}
%The divergence of the above EMT along the world lines of the volume elements reads:
%\begin{equation}
%u_{\mu}\Delta {T^{\mu\nu}_{(i);\nu}}= P_{ci} \Theta,
%\end{equation}
%where $\theta =3H$ is the scalar of expansion, whereas 
with the energy conservation law for each component becoming
\begin{equation}\label{ECL2}
u_{\mu}{T^{\mu\nu}_{(i){;\nu}}} = \dot{\rho_i} + \Theta(\rho_i + p_i + P_{ci}) = 0, 
\end{equation}
where $\Theta=3H$ is the scalar of expansion. Note also that if $P_{ci}$ is negligible (no particle production), the equilibrium energy conservation law  is recovered. Such a condition will be more physically defined below. 

In the presence of a gravitational particle source, the balance equation for the particle flux and entropy current are redefined in order to derive the creation pressure of each component.  The particle flux is $N^\mu_{(i)}=n_iu^{\mu}$, and its divergence takes the form
\begin{equation}\label{PCL2}
 {N^\mu_{(i)}};_{\mu} \equiv \dot{n_i} + n_i \Theta= n_i \Gamma_{iN} \,\,\,\Leftrightarrow\,\,\,\frac{\dot N_i}{N_i} = \Gamma_{iN},
\end{equation}
where by definition the total number of particles in the comoving volume, $N_i = n_i a^{3}$, and $[\Gamma_{iN}]$ with dimension of $[time]^{-1}$ is the number particle creation rate of the $i$-th component. Naturally, when compared with $\Theta$ this new microscopic time scale quantifies the efficiency of the gravitational particle production. In particular, if $\Gamma_{iN} << \Theta$, the creation process according to (\ref{PCL2}) can safely be neglected.  

In the same vein, the  entropy current reads:
\begin{equation}\label{EC}
S_{(i)}^\mu = s_i u^\mu \equiv n_i \sigma_i u^\mu,
\end{equation}
where $\sigma_i$ is the specific entropy per particle. Now, by taking the 4-divergence of the first equality above, the irreversible creation process implies that:
\begin{equation}\label{DivS1a}
 {S^\mu_{(i);\mu}} =  \dot s_i +  s_i \Theta=s_i\Gamma_{iS}\,\,\,\Leftrightarrow\,\,\,\frac{\dot S_i}{S_i} = \Gamma_{iS},
 \end{equation}
 where we have also defined $S_i = s_ia^{3}$ and $\Gamma_{iS}$ is the entropy creation rate, a new (irreversible) time scale (dimension $[\Gamma_{iS}]\equiv [time]^{-1}$).  In addition, $\Gamma_{iS} \geq 0$ because the constraint defining  the second law of thermodynamics must be satisfied. 
 
It is worth noticing the difference between equations (\ref{PCL2}) and (\ref{DivS1a}).  The first one is clearly related with the emergence of particles in the spacetime which must also affect the fluid entropy production. Under certain conditions that will be  discussed below, the existence of this second time scale implies that the  variation rate of the specific entropy may be different from zero. In fact, since $\sigma_i = S_i/N_i$, its time-comoving  derivative combined with (\ref{PCL2}) and (\ref{DivS1a}) yields 
\begin{equation}\label{dotsigma}
\dot \sigma_i = \sigma_i(\Gamma_{iS} - \Gamma_{iN}). 
\end{equation}
 Hence, $\dot\sigma_i = 0$ only for two different situations: (i) equilibrium states when $\Gamma_{iS}=\Gamma_{iN}\equiv 0$, and (ii) nonequilibrium states ($\dot S_i,\dot N_i \neq 0$), but $\Delta \Gamma_i= 0$ so that $\Gamma_{iS}=\Gamma_{iN}$. Following the nomenclature introduced long ago for a one-component fluid \cite{CLW92,LG92}, this case it will also dubbed here  ``adiabatic" creation  and will be discussed separately (see subsection IIIC below). As we shall see later, $\Gamma_{iS}\geq \Gamma_{iN}$. Hence, in general $\dot \sigma_i \geq 0$ for matter creation models with the specific entropy produced coming from the ``uncompensated heat" spent for thermalization of the created particles for any decoupled component.  It is also worth notice that for $\Gamma_{iN}\equiv 0$ but $\Gamma_{iS}\neq 0$,  we are describing (for each component) the pure phenomenon of bulk viscosity due to the universe expansion. In this case, $\dot \sigma_i > 0$ and the CMB thermal spectrum is destroyed in the course of the expansion. However, this does not happens in the ``adiabatic" case (see section III).  

At this point, one may  ask: \textit{What about the creation pressure and the temperature law for this general case?} Such topics, including the ``adiabatic case", will be separately discussed  in the next subsections. 

\subsection{Creation Pressure}

To begin with, we first remark that even in the presence of a dissipative processes like matter creation,  the local equilibrium hypothesis means that each component satisfies the local form of the Euler relation \cite{callen} 
the comoving time derivative of the Gibbs law  
\begin{equation} \label{entropyd}
n_i T_i \dot \sigma_i= d\rho_i - (\frac{\rho_i + p_i}{n_i})dn_i,
\end{equation}
 Now, by taking the comoving time derivative of the above expression, and combining the result with the balance equation for the particle number density we find:

%y using the  Gibbs law, the above expression leads (for each component) to the so-called Gibbs-Duhem relation: 
%\begin{equation}\label{GB}
%n_i\sigma_i dT_i = dp_i - n_id\mu_i. 
%\end{equation}
%It means the existence of only two independent thermodynamic variables, say, $n_i$ and $T_i$, and, as such, one may assume  %that  $\rho_i=\rho_i(T_i,n_i)$ and $p_i =p_i(T_i,n_i)$. Now, combining the energy conservation laws with  the thermodynamic %identity: 
\begin{equation}\label{dotsigma1}
n_i\,T_i\,\dot {\sigma_i} = \dot \rho_i +  (\rho_i+p_i)\Theta - (\rho_i + p_i)\Gamma_{iN}. 
\end{equation}
Thus, in order to obtain the general form of the creation pressure, it is enough to consider the energy conservation law for each component (\ref{ECL2}) plus the variation rate of  the specific entropy (\ref{dotsigma}). The general form of the creation pressure reads:
\begin{equation}\label{CP}
P_{ci} =  - (\rho_i + p_i) \frac{\Gamma_{iN}}{\Theta} - \frac{n_i \sigma_i T_i}{\Theta} (\Gamma_{iS} - \Gamma_{iN}). 
\end{equation}
Ultimately, the above quantity must be incorporated in the Einstein field equations endowed with creation of all components. It is immediate to see that the set of  independent FLRW equations take the irreversible form:

\begin{eqnarray}
%\begin{split}
&&\rho_{\text{T}}\equiv\sum_{i=1}^{N} \rho_i =3M^{2}_P\,H^{2}\,,\label{EE1b}\\ 
&&p_{\text{T}} \equiv \sum_{i=1}^{N} (p_i + P_{ci}) =-M^2_{P} \left[2\dot H+3H^{2}\right]\label{EE2b}\,,
%\end{split}
\end{eqnarray}
where $\rho_{\text{T}}$ and $p_{\text{T}}$ are, respectively, the total energy density and pressure while $M_\text{P}=(8\pi \text{G})^{-1/2}\simeq 2.4\times 10^{18} \, \text{GeV}$ is the reduced Planck mass, and  $P_{ci}$ is given by (\ref{CP}). Note that the  late time $\Lambda$CDM model is readily recovered by assuming $P_{ci}=0$, $N=5$ and a matter-energy content formed by (1) CDM, (2) baryons, (3) radiation, (4) neutrinos, and  (5) a dark energy represented by a vacuum state with negative pressure, $p_v=-\rho_{v}$. As usual, the model  is completed when such decoupled components are described  by the equation of state (EoS),  $p_i = \omega_i \rho_i$, where $\omega_i$ is contained in the interval $[-1,1]$. 

Note also that creation pressure above is always negative. In this case  one may choose $\Omega_{DE}=0$ and, as such, this kind of scenario may provide a description of the present accelerating stage of the Universe without vacuum energy density, thereby reducing the dark sector. In single-fluid description (CCDM), some  possibilities have been discussed in the literature (see introduction).
 
 %As occurs with the bulk viscosity and other dissipative mechanisms in the first order thermodynamics, the Friedmann equation for the energy density (\ref{EE1b}) is not modified. Note also that since $\Gamma_{iS} \geq \Gamma_{iN}$, 

\subsection{Temperature Law}

Let us now discuss how  the temperature evolution law is modified by the particle creation  ($\Gamma_{iN}$) and entropy production ($\Gamma_{iS}$) rates.
%The procedure is a natural generalisation from that one applied for equilibrium processes [see Eq. (\ref{TI})]. 

By taking the pair ($n_i,T_i$) as dependent variables, using the balance equation for the particle number density (\ref{PCL2})  and the thermodynamic identity:
\begin{equation}\label{TI}
T_i \biggl(\frac{\partial p_i}{\partial T_i}\biggr)_{n_i}=\rho_i + p_i - n_i
\biggl(\frac{\partial \rho_i}{\partial n_i}\biggr)_{T_i},
\end{equation}
it is readily checked that the variation rate of the temperature takes the form:
\begin{equation}
    \left(\frac{\partial \rho_i}{\partial T_i}\right)_{n_i}{\dot{T_i}} = -T_i\left(\frac{\partial p_i}{\partial T_i}\right)_{n_i}\,\frac{\dot n_i}{n_i} - P_{ci}\Theta - (\rho_i + p_i)\Gamma_{iN}\,,
\end{equation} 
or still, by inserting the second equality of (\ref{PCL2}) and  the creation pressure (\ref{CP}): 
%\begin{equation}
%\frac{\dot{T_i}}{T_i} = \left(\frac{\partial p_i}{\partial \rho_i}\right)_{n_i}\,\frac{\dot n_i}{n_i} %- \frac{P_{ci}\Theta}{T_i} - (\frac{\rho_i + p_i}{T_i})\Gamma_{iN}\,,
%\end{equation} 
%or still, inserting expression (\ref{CP}) for the creation pressure 
\begin{equation}\label{TF}
\frac{\dot{T_i}}{T_i} = -\left(\frac{\partial p_i }{\partial \rho_i} \right)_{n_i} [\Theta- \Gamma_i] + \frac{n_i\sigma_i (\Gamma_{iS} - \Gamma_{iN})} {(\partial \rho_i/\partial T_i)_{n_i}}\,.
\end{equation}
Note that whether $\Gamma_{iS}=\Gamma_{iN} \neq 0$, the form of the equilibrium temperature law
\begin{equation}\label{Templaw1}
\frac{\dot{T_i}}{T_i} = \left(\frac{\partial p_i}{\partial \rho_i}\right)_{n_i}\,\frac{\dot n_i}{n_i},    
\end{equation}
is readily recovered, as should be expected. For $\Gamma_{iS} \neq \Gamma_{iN}$, we obtain the temperature law for general ``nonadiabatic" case, since $\dot\sigma_i\neq 0$ [see (\ref{dotsigma})]. Now, due to its physical importance, the\, ``adiabatic" case, that is,  $\Gamma_{iS} = \Gamma_{iN}\equiv \Gamma_{i}$ will be separately discussed next. 

\subsection{The ``Adiabatic" Case}
 The ``adiabatic'' creation  in the decoupled multi-fluid description is defined by $\dot{\sigma}_{(i)} = 0$, that is, $\Gamma_{iS} = \Gamma_{iN}\equiv\Gamma_{i}$ [see discussion below Eq. (\ref{dotsigma})]. In this case the creation pressure (\ref{CP}) becomes
\begin{equation}\label{Pci}
P_{ci} = -(\rho_{i} + p_{i})\frac{\Gamma_{i}}{\Theta} = - (\rho_{i} + p_{i})\frac{\Gamma_{i}}{3H}, 
\end{equation}
where $\Gamma_{i}$ is positive definite because the second law of thermodynamics. Actually, in  this case the balance  
equation for the entropy boils down to:
\begin{equation}\label{DivS1}
{S^\mu}_{(i);\mu} = {\dot s_i} + s_i\Theta = s_i\Gamma_i \geq 0 
\end{equation}
Hence,  one may conclude from (\ref{PCL2}) that the Universe may only create matter (${\dot N_i}>0$). In addition,  for the expanding Universe ($ H > 0$), the associated creation pressure of each decoupled component is always  negative. This generalises the original results of Prigogine et al. \cite{P89} for a single-fluid approach  with irreversible particle creation. Of course,  it also explains why a generic multi-fluid cosmology may accelerate at low redshifts mimicking (for non-relativistic components) the $\Lambda$CDM model (see introduction).

Further, since  $\sigma_i = S_i/N_i$, where $S_i = s_i a^{3}$  is the entropy in a comoving volume, and $N_i = n_i a^{3}$, the condition  $\dot{\sigma_i} = 0$ also implies that 
\begin{equation}\label{SN}
\dot S_i/S_i =\dot N_i/N_i\,\,\,\,  \Leftrightarrow\,\,\, S_i = k_B N_i.
\end{equation}
Therefore, the entropy growth associated to this gravitational particle production process is actually closely related with the quantum emergence of particles in the space-time. As we shall see, the created particles  are in thermal equilibrium with the existing ones.  An important point to keep in mind here is that the presence of the creation pressure in this macroscopic description is not the result of a collisional process as happens, for instance, with the standard bulk viscosity mechanism.  
%\begin{equation}\label{TEMP}
%\frac{\dot{T_i}}{T_i}=-\left(\frac{\partial P_i }{\partial \rho_i} \right)_{n_i} \frac{\dot n_i}{n_i}-\frac{P_{ci}\Theta + n_i \Gamma_i (\partial \rho_i / \partial n_i)_{T_i}}{T_i(\partial \rho_i / \partial T_i)_{n_i}}.
%\end{equation}

In the ``adiabatic" case the temperature law  is also directly obtained from  (\ref{TF}) by taking   $\Gamma_{iS} = \Gamma_{iN} = \Gamma_i$. Hence, the temperature law (\ref{TF}) reduces to  
\begin{equation}\label{Templaw} 
 \frac{\dot{T_i}}{T_i} = -\left(\frac{\partial p_i}{\partial \rho_i}\right)_{n_i} \left(\Theta - \Gamma_i\right)=\left(\frac{\partial p_i}{\partial \rho_i}\right)_{n_i}\,\frac{\dot n_i}{n_i},
\end{equation}
thereby recovering the standard equilibrium relation in the limit $\Gamma_i \rightarrow 0$. In the nonrelativistic approximation \cite{Pauli}, the EoS reads
\begin{equation}\label{EoS2}
\rho_i=n_i m_i + \frac{3}{2} n_iT_i= n_im_i + \frac{3}{2}p_i,
 \end{equation}
with  the temperature law (\ref{Templaw}) assuming the modified form:
\begin{equation}\label{temp2}
\frac{\dot{T_i}}{T_i}= -2\frac{\dot a}{a}+\frac{2}{3} \Gamma_i, 
\end{equation}
which for $\Gamma_i = 0$ also reduces to the standard equilibrium result, $T_i \propto a^{-2}$.  By using the EoS, $p_i=\omega_i\rho_i$, we also see that the temperature law  for arbitrary values of $\omega_i$ becomes: 

\begin{equation}\label{temp2a}
 \frac{\dot{T_i}}{T_i}=-3\omega_i\frac{\dot a}{a}+\omega_i\Gamma_i.   
\end{equation}  
Hence, for $\Gamma_i = 3 \beta_i H$, a simple integration of the above equation yields, $T_i=T_{0i}(1+z)^{3\omega_i(1-\beta_i)}$. In particular, for radiation (CMB)  $\omega_i=1/3$,  $\beta_i\equiv \beta$, this expression  reduces to $T=T_{0}(1+z)^{1-\beta}$, where $\beta$ would be determined by the astronomical observations \cite{Lima97}. Later on, several authors investigated how $\beta$  would be constrained by using  the absorption lines of quasars at the redshift of the absorber \cite{TempL,Mol2002}, as well as from Sunyaev-Zeldovich effect \cite{LoSecco2001,Cui2005,Luzzi2009,Jetzer2011,Norte2011,Muller2013,Hurier2014,KK2015,Ave2016,BJL2019}. More recently, other relations for ``adiabatic" production based on different phenomenological expressions for $\Gamma_i$ have also been proposed and constrained by the existing observations \cite{nunes2016,BJL2019,Cardenas2020,LTS2021,M2021}. Note also that for CMB or more generally for massless (bosonic or fermionic) particles ($\omega_i=1/3$), the temperature evolution (\ref{temp2a}) can be rewritten as 
\begin{equation}\label{temp3}
 \frac{\dot T_i}{T_i}=-\frac{\dot a}{a}+\frac{\Gamma_{i}}{3} \leftrightarrow  \frac{\dot T_i}{T_i}= -\frac{\dot a}{a}+\frac{\dot N_i}{3N_i}, 
\end{equation} 
and a simple integration yields:
\begin{equation}\label{temp4}
T_i{(t)}a(t)N_i(t)^{-1/3}= const. \rightarrow T_i=T_{0}(1+z)(\frac{N_i(t)}{N_0})^{1/3}.
\end{equation}
Therefore, if the average number of photons $N_i(t)$ is constant (no photon creation), the standard CMB temperature law is recovered. In addition, since $N(t)\leq N_0$, it follows  that the value of the temperature for a finite redshift is always smaller than the one predicted by the $\Lambda$CDM model. 

Now, some comments are in line in order to stress the generality of the above temperature law. Firstly, it is widely believed that cosmological creation of photons in the expanding Universe is not allowed  because the blackbody form of the CMB spectrum is destroyed \cite{W72,Steigman78,KT90,DOD2003}. However, as discussed long ago and rigorously proved recently \cite{LTS2021}, the blackbody form of the CMB spectrum is still preserved when gravitational photon production occurs under ``adiabatic" conditions. There is a twofold reason for that: (i) the second equality in (\ref{Templaw}) has the same equilibrium form. In fact, for $\omega_i =1/3$, simple integration yields $n_i \propto T^{3}_i$. In addition,  from $\dot{\sigma_i}=0$, the Gibbs law (\ref{dotsigma1}) yields  $\rho_i \propto n^{4/3}_i$ so that $\rho_i \propto T^{4}_i$, which are the same equilibrium relations for blackbody radiation,  and (ii) as we shall see next section, the preservation of the equilibrium shape for any component in the mixture is also a direct consequence of the modified relativistic Boltzmann equation recently derived by incorporating matter creation under ``adiabatic" conditions. This is an interesting point because a more detailed study of the CMB anisotropies and distortions  requires the previous knowledge of the conditions under which a  blackbody shape is preserved.  

\section{Modified Boltzmann Equation and Particle Creation}
% New temperature law and CMB blackbody spectrum}

From a kinetic viewpoint, the behavior of a decoupled multi-fluid mixture can properly be derived  by following  the evolution of each  phase space density, $f_{(i)}(x^{\mu}_{(i)},P^{\mu}_{(i)})$.  If one includes  gravitational matter creation of all components, this of course must be governed by a suitable modification of the Boltzmann equation.  

Let us first recall that in the relativistic kinetic framework, the basic macroscopic quantities (fluxes) are microscopically defined taking the averaging over the distribution function\cite{KT90,JL13}. For each component we have:
 %\cite{LB2014,BL2015}.
\begin{eqnarray}
    \label{KF1}
    N_{(i)}^{\mu}&=& \frac{g_{(i)}}{(2\pi)^3}\int f_{(i)}\,{P^\mu_{(i)}}\sqrt{g}\frac{d^3P_{(i)}}{P_{(i)}^0}, \\
    \label{KF2}
    S_{(i)}^\mu&=& -\frac{g_{(i)}}{(2\pi)^3}\int\left[f_{(i)} \ln f_{(i)} - f_{(i)}\right]P^\mu_{(i)}\sqrt{g}\frac{d^3P_{(i)}}{P^0_{(i)}}, \\
    \label{KF3}
    T_{(i)}^{\mu \nu}&=& \frac{g_{(i)}}{(2\pi)^{3}}\int{f_{(i)} \, P_{(i)}^\mu P_{(i)}^\nu \sqrt{g}\frac{d^3P_{(i)}}{P^0_{(i)}}}, 
\end{eqnarray}
where $g_{(i)}$ counts the internal degrees of freedom (spin states degeneracy) of a given component, g is the metric determinant, $P^{\mu}$ is the comoving momentum, and, as before, the indexes ({\it i}) denotes the $\textit{i-th}$ component in the mixture. Henceforth, unless explicitly stated, all repeated Latin scripts in round brackets are not summed. 

The equilibrium states associated to the $i-$th decoupled component in the mixture is described by a distribution function. For a relativistic non-quantum weakly-interacting dilute gas, it  assumes the form, $f_{(i)} = exp{(\alpha_i - \beta_i E_i)}$, where $\alpha_i(t) = \mu_i/T$ defines the relativistic chemical potential and $\beta_i(t)$ is the inverse of temperature \cite{JB88}.  Such  a form is a solution of the standard collisionless Boltzmann equation (CBE). It is widely known that when the mass shell condition, $g_{\mu\nu}{P^\mu}_{(i)} {P^\nu}_{(i)}\equiv m_{(i)}^2$, is imposed for the physical momentum in a flat geometry [$p_{(i)} = a(t)P_{(i)}$, $f_{(i)}=f_{(i)}(t,p_{(i)})$], the CBE can be written as \cite{KT90,JB88,JL13} 

\begin{equation}
    \label{MSCBE}
        \frac{1}{E_i}\mathcal{L}[f_{(i)}]\equiv \frac{\partial f_{(i)}(t,p_{(i)})}{\partial t} - H p_{(i)} \frac{\partial f_{(i)}(t,p_{(i)})}{\partial p_{(i)}} = 0 
\end{equation} 
%C (f_i,f'_i)
where ${\mathcal{L}}[f_{(i)}]$ is the standard Liouville operator.

It is readily verified using (\ref{MSCBE}) that the kinetic definitions (\ref{KF1}) - (\ref{KF3})  reproduce the {\it macroscopic equilibrium expressions}  for $N^{\mu}_{(i)}, S_{(i)}^{\mu}, T_{(i)}^{\mu \nu}$ and also the equilibrium  conservation equations in the FLRW metric, namely:  $N_{(i)}^{\mu};_{\mu}=0,  S_{(i)}^{\mu};_{\mu}=0$, and also the energy conservation law, $u_\mu T_{(i)}^{\mu \nu};_{\nu}=0$ (see calculations in  \cite{JB88,JL13}). 
%These standard results will be recovered as a particular case of the ``adiabatic" creation kinetic formalism.

At this point, it is also natural to ask: {\it What happens in the presence of gravitationally induced matter creation?}  In the next two sections, it will be shown that an appropriated  (collisionless) modified Boltzmann equation ({\bf MBE}) also reproduce all the results of section III when gravitational matter creation occurs under ``adiabatic" conditions (see also appendix A).  Only in this case, the equilibrium shape of the distribution function is preserved both for massive and massless particles.

To begin with, let us  first recall that the distribution function, $f_{(i)}(x^{\mu}_{(i)}, P^{\mu}_{(i)})$, for each component, must be a solution of the MBE. In the present context, a basic requisite is that particle and entropy productions must be included (see \cite{LB2014} for a single fluid), in agreement with the second equation in (\ref{SN}). For a decoupled self-gravitating mixture, its manifestly covariant expression in terms of the comoving momentum takes the form:
%follows as an extension of the one fluid expression
%\begin{equation}\label{Beq1}
%p^\mu_{(i)}\frac{\partial{f_{(i)}}}{\partial{x^\mu_{(i)}}}-\Gamma^{\mu}_{\alpha %\beta}p^\alpha_{(i)}p^\beta_{(i)}\frac{\partial f_{(i)}}{\partial p^\mu_{(i)}} - %{\mathcal P_{gi}}(x^{\mu}_{(i)},p^{\mu}_{(i)}) = C (f_i,f'_i)\,,
%\end{equation}
\begin{equation}\label{Beq1}
P^\mu_{(i)}\frac{\partial{f_{(i)}}}{\partial{x^\mu_{(i)}}}-\Gamma^{\mu}_{\alpha \beta}P^\alpha_{(i)}P^\beta_{(i)}\frac{\partial f_{(i)}}{\partial p^\mu_{(i)}} + {\mathcal P_{Gi}}(x^{\mu}_{(i)},P^{\mu}_{(i)}) = C [f_{(i)},f'_{(i)}]\,,
\end{equation}
where  $P^{\mu}_{(j)} \equiv (E_{(j)}, P^{i}_{(j)})$ is the four-momentum and the geodesic equation has been used to rewrite the second term. $C[f_{(i)},f'_{(i)}]$ is the standard collisional term, including all possible interactions and sources of distortions  changing the form of the equilibrium distribution.  For cosmic background radiation (CMB), for instance, it also includes Compton scattering, double Compton and Bremsstralung emissions. The physical consequences of such a term have already been quite explored in the literature \cite{KT90,DOD2003} but, for a while, we consider $C[f_{(i)},f'_{(i)}]\equiv 0$ thereby focusing our attention over the ``adiabatic" gravitationally induced creation contribution.  

The extra term in the left hand side ({\it l.h.s.}) of (\ref{Beq1}), ${\mathcal P}_{Gi} (x^{\mu}_{(i)},P^{\mu}_{(i)})$, is a non-collisional source term describing the gravitationally induced particle production process due to the expansion of the Universe. This process cannot be thought as a kind of particle injection whose momentum must be lately redistributed thereby repopulating the distribution, and, as such, provoking distortions in the equilibrium spectrum. Although being responsible for a dynamical creation pressure and modifying the temperature law, it does not change the equilibrium shape of the distribution function.  This explain why it was written in the left hand side of the above modified Boltzmann equation. Its choice is dictated by two simple criteria \cite{LB2014,LTS2021}: (i) ${\mathcal P}_{Gi} \propto \Gamma^i_{\lambda \nu}$ since ${\mathcal P}_{Gi} [f_{(i)}]$ should disappear in the absence of gravity when the Levi-Civita connections are  identically null  ($g_{\mu\nu}=\eta_{\mu\nu}$), and (ii) ${\mathcal P}_{Gi} \propto  {\Gamma_{i}}/{\Theta}$.  Such a condition is suggested by the macroscopic equations [see ($\ref{PCL2}$) and ($\ref{DivS1a}$)] when  $\dot\sigma_i=0$\,(``adiabatic" case).  As in the macroscopic approach, $\Gamma_{i}$ represents the creation rate of the $\textit{i-th}$ fluid component in the mixture. 

Now, it is worth noticing that the constraint derived from the ``mass shell" condition, $g_{\mu\nu}{P^\mu}_{(i)} {P^\nu}_{(i)}\equiv m_{(i)}^2$, has not been imposed in the above expression. By neglecting the standard collisional term, $C[f_{(i)},f'_{(i)}] \equiv 0$, the mass shell Boltzmann equation in terms of the local momentum can be written as:
%($p^i=aP^{i}$)
\begin{equation}
    \label{MSBE}
\frac{1}{E_i}{\mathcal L}[f_{(i)}]\equiv \frac{\partial f_{(i)}}{\partial t} - H p_{(i)} \frac{\partial f_{(i)}}{\partial p_{(i)}}  + \frac{\Gamma_{i}}{3} p_{(i)} \frac{\partial f_{(i)}}{\partial p_{(i)}} = 0.
\end{equation} 
where $f_{(i)}=f_{(i)}(t,p_{(i)})$, where $p_{(i)}$ is the modulus of the momentum of the {\it i-th} decoupled component (for more details see Appendix A).

Note that  the non-null creation rates $\Gamma_{i}$ which is a consequence of the expanding Universe, contributes at the level of the Liouville operator like the Hubble parameter (with a changed sign), as should be physically expected for a purely gravitational effect. In addition, for $\Gamma_{i} << 3H$ such a term is negligible as previously determined based on the macroscopic approach (see section III) thereby  reducing (\ref{MSBE}) to the standard collisionless Boltzmann equation without creation. In what follows the above equation will be justified by deriving kinetically the macroscopic balance equations  with ``adiabatic" creation (see also Appendix A for more details).

\section{Kinetics versus Thermodynamics: Recovering the Macroscopic Results with ``Adiabatic" Creation}
%Energy-Momentum, Creation Pressure and Temperature} 

Let us now show how the modified Boltzmann equation
(50) allow us to recover the basic macroscopic balance equations 
for the particle and entropy fluxes, as well as, 
the energy-momentum tensor including the creation
pressure. All the derivations are based on the kinetic
definitions for an arbitrary number of decoupled
components, as given in the previous section [see (\ref{KF1})-(\ref{KF3})]. 

\subsection {Particle Flux}

To begin with, let us combine the flat FLRW geometry (\ref{FRW}) and (\ref {KF1}). As one may check, the only non-null component of the particle flux is the particle number density itself:  
\begin{equation}
N_{i}^{0}=n_i= \int f_{(i)} d^3 p_{(i)}.
\end{equation}
Hence,  the divergence of the particle flux becomes:
\begin{equation}
N^\mu_{(i);\mu} = \frac{1}{a^3} \frac{\partial}{\partial t}\left(a^3 \int f_{(i)} d^3 p_{(i)}\right),
\end{equation}
and by expanding the product and using the collisionless Boltzman equation (\ref{MSBE}) we find: 
\begin{equation}\label{NK}
N^\mu_{(i);\mu}= 3Hn_{i} + H \left(1 - \frac{\Gamma_{i}}{3H}\right) \int^{\infty}_0 p_{(i)} \frac{\partial f_{(i)}}{\partial p_{(i)}} d^3p_{(i)}.
\end{equation}
Now, integrating by parts and using that for a well behaved distribution, the product $p^{3}f(p)$  vanishes in the limits of integration, one finds:
\begin{equation*}
\int^{\infty}_0 p_{(i)} \frac{\partial f_{(i)}}{\partial p_{(i)}} d^3 p_{(i)} = -3n_{i},
\end{equation*}
 and inserting the above result into (\ref{NK}), we recover the  macroscopic evolution equation for the particle number density  with ``adiabatic" creation [cf. Eq. (\ref{PCL2})]
\begin{equation}\label{KNB}
N^\mu_{(i);\mu} = n_i \Gamma_{i}.
\end{equation}

\subsection {Entropy Flux}

Similarly, the balance equation for $S_{(i)}^{\mu}$ may be derived based on the previous approach for $N_{(i)}^{\mu}$. The only non-null component of the entropy flux $S_{(i)}^{\mu}$ also defines the entropy density [see kinetic definition in (\ref{KF2})] 
\begin{equation} 
S_{(i)}^{0}= s_i = -\int [f_{(i)} \ln f_{(i)} - f_{(i)}] d^3 p_{(i)},
\end{equation}
while the divergence of the entropy flux reads:
\begin{align}
S^\mu_{(i);\mu} = -\frac{1}{a^3} \frac{\partial}{\partial t}\left(a^3 \int \left[f_{(i)} \ln f_{(i)} - f_{(i)} \right] d^3 p_{(i)}\right).
\end{align}
Then, by expanding the product, and again using the collisionless Boltzman equation with creation (\ref{MSBE}) we find: 
\begin{equation}\label{SF}
S^\mu_{(i);\mu} = 3Hs_{(i)} - H \left(1 - \frac{\Gamma_{i}}{3H}\right) \int p_{(i)} \frac{\partial f_{(i)}}{\partial p} \ln f_{(i)} d^3 p_{(i)}.
\end{equation}
To proceed further we need to solve the integral in the last term. By adopting spherical coordinates and solving it by parts  we obtain   %switching to the spherical coordinate system; and considering the validity of the relationship
\begin{equation*}
\int^{\infty}_0 p_{(i)} \frac{\partial f_{(i)}}{\partial p_{(i)}} \ln f_{(i)} d^3 p_{(i)} = 3s_i,
\end{equation*}
where it was used that the product $p_{(i)}^{3}[f_{(i)}\ln f_{(i)} - f_{(i)}]$  vanishes in the integration limits. Now, by inserting the above result into (\ref{SF}), the kinetic balance equation for the entropy  of a decoupled mixture with creation 
\begin{equation}\label{KSB}
S^\mu_{(i);\mu} = s_i \Gamma_i,
\end{equation}
is recovered. This result is clearly a consequence of the modified Boltzmann equation being also in perfect agreement with (\ref{DivS1a}) appearing in the macroscopic approach when ``adiabatic" conditions are assumed ($\Gamma_{iS}=\Gamma_{iN}=\Gamma_i$). 

\subsection {Creation Pressure and Energy-Momentum Tensor}

In order to obtain kinetically the creation pressure, let us multiply by $E_i$ the modified Boltzmann equation (\ref{MSBE}). Now, by integrating the result over the momentum space, it follows that 

\begin{equation}
    \label{MSBE2}
\int^{\infty}_0 E_{(i)}\frac{\partial f_{(i)}}{\partial t} d^3 p_{(i)} - H \left(1- \frac{\Gamma_{i}}{3H}\right) \int^{\infty}_0 E_{(i)} p_{(i)} \frac{\partial f_i}{\partial p_{(i)}}d^3 p_{(i)}=0.  
\end{equation} 
A simple  integration term by term yields:
\begin{equation}\label{ECLK}
\dot{\rho_i} + \Theta(\rho_i + p_i) - (\rho_i + p_i)\Gamma_{i} = 0,
\end{equation}
which can be rewritten as:
\begin{equation}\label{PCiK}
\dot{\rho_i} + (\rho_i + p_i + P_{ci})\Theta = 0, \,\,\Leftrightarrow \,\, P_{ci}= -(\rho_i + p_i)\frac{\Gamma_{i}}{\Theta}.
\end{equation}
As should be expected, the creation pressure above is exactly the same macroscopic expression for the ``adiabatic" case [see Eq. (\ref{Pci})]. Therefore,  the creation rate $\Gamma_{i}$ for each decoupled component appearing in the modified Boltzmann equation (\ref{MSBE}), also modulates the noncollisional correction term that disappears in the special relativistic limit when $\Gamma^{\alpha}_{\beta \gamma} \equiv 0$. 

Naturally, the above result implies that the energy conservation law can also be obtained from the kinetic definition of the EMT given by (\ref{KF3}). In order to show that let us now calculate the divergence of the total energy-momentum tensor projected onto the four-velocity $u_\mu$. 
%It involves a rather delicate and subtle aspect from a kinetic viewpoint since the Einstein field equations requires a divergenceless  energy-momentum tensor and we know that the creation pressure has a non-collisional origin. In addition, like the bulk viscosity, we are aware that that the creation pressure is also negative for an expanding Universe [see Eq. (\ref{Pci}].  
Firstly, it should be recalled that unlike to what happens with the energy density, there is no constraint conditions to the kinetic pressure for states out of equilibrium \cite{JB88}. Thus,  it is also natural to assume  the existence of a  corrective   (non-collisional) creation pressure term. Now, let us also assume that homogeneity and isotropy dictates the following form $\Delta {T^{l}_{(i)k}}  = -{ P_{ci}}\delta^{l}_k$, or equivalently, ${\Delta {T^{\mu \nu}}_{(i)}}= -{P_{ci}}h^{\mu \nu}$. However, for the sake of generality, we consider for a while that  $P_{ci}$ is an {\it unknown} creation pressure not necessarily equal to the value given by (\ref{PCiK}). Thus,  the total EMT for each component is $T^{\mu \nu}_{(i)} = T^{\mu \nu}_{(i)|E}  + \Delta {T^{\mu \nu}_{(i)}}$, being kinetically defined by the  expression  (\ref{KF3}). In this case, we  can write for the projected divergence:
\begin{equation}
u_\mu T^{\mu \nu}_{{(i)};\,\nu}\equiv u_\mu \left[ \frac{1}{a^3}\frac{\partial }{\partial x^\nu}(a^3T^{\mu \nu}_{(i)})+\Gamma^{\mu}_{\alpha \beta}T^{\alpha \beta}_{(i)}\right],
\end{equation}
 Now,  by summing over the repeated indices and using the last expression in (\ref{Csymb}) it becomes: 
\begin{eqnarray}
u_\mu T^{\mu \nu}_{{(i)};\,\nu}&\equiv&\frac{1}{a^3}\frac{\partial}{\partial t}(a^3T^{00}_{(i)|E})+\Gamma^0_{ij}(T^{ij}_{(i)|E} + \Delta {T^{ij}_{(i)})} \nonumber \\
&=&\frac{1}{a^3}\frac{\partial f_i}{\partial t}\left(a^3\int^{\infty}_0 {f_i E_i} d^3p_{(i)}\right)+3\frac{\dot a}{a} (p_i + P_{ci}), \nonumber \\
\end{eqnarray}
or equivalently, 
\begin{equation}
 u_\mu T^{\mu \nu}_{{(i)};\,\nu} = 3H(\rho_i + p_i + P_{ci}) + H\left(1-\frac{\Gamma_{i}}{\Theta} \right)\int^{\infty}_0 E_i p_{(i)}\frac{\partial f_{(i)}}{\partial p_{(i)}} d^3p_{(i)},
 \end{equation}
 and solving the integral by parts we find
 \begin{eqnarray}
  u_\mu T^{\mu \nu}_{{(i)};\,\nu}&=& 3H(\rho_i + p_i + {P_{ci}}) - 3H\left(1-\frac{\Gamma_i}{\Theta}\right)(\rho_i + p_i), \nonumber \\
&=& 3HP_{ci} + (\rho_{i} + p_{i})\Gamma_{i} = 0.
\end{eqnarray}

\hspace{0.1cm}Therefore,  as  required by the Einstein gravitational equations, the projected divergenceless total energy momentum-tensor ($u_\mu T^{\mu \nu}_{(i);\,\nu}=0$), that is, the expression of the energy conservation in the FLRW geometry, is obtained only when the creation pressure ${P_{ci}}$ is given by the previously derived expression [see the second expression in Eq. (\ref{PCiK})]. 

\subsection{Temperature Evolution Law}

Let us now proceed to calculate the temperature evolution for the decoupled fluid mixture endowed with ``adiabatic" gravitational particle production based on our extended kinetic approach. In this section we assume that the distribution function for a non-quantum relativistic gas endowed with  ``adiabatic" creation is also given by the standard equilibrium form:
    \begin{equation}\label{EDF}
        f_{(i)} = {e^{\alpha_i (t)-\beta_i{(t)} E_i}},
    \end{equation} 
    where as before $\alpha_i$ is a scalar function and $\beta_i (t)$ can be interpreted as the inverse of temperature [see discussion right before Eq. (\ref{MSCBE})].
    %\textbf{(note once again that the thermal equilibrium mixture is being taken into account)}. 
    
    The main aim here is to show kinetically that such a  form is preserved  if and only if the temperature evolution law is modified in agreement with the general macroscopic law and a generic creation rate $\Gamma_{i}$ [see Eq. (\ref{temp2a})]. 
    
    Now, by inserting  (\ref{EDF}) into the modified Boltzmann's equation (50) we obtain: 
    \begin{equation}\label{TKG}
         \dot \alpha_i -\dot \beta_i E_i+ \beta_i H\left( 1-\frac{\Gamma_{i}}{\Theta} \right)\frac{p^2}{E_i}=0.
    \end{equation}
    This equation has two extreme limits. The  nonrelativistic limit ($m_i \gg T$, $E_i \simeq m_i +\frac{p^2}{2m_i}$) and the ultrarelativistic or negligible rest mass limit  ($m_i << T, E_i \simeq p_{(i)}$). Let us now determine the solutions for such limits separately. 
    
    \begin{itemize}
    
    \item The non-relativistic domain ($m_i >> T_i$)
    \vskip 0.1cm
    In this limit, the above  equation (\ref{TKG}) takes the form:
    \begin{equation}
         \frac{\dot \alpha_i}{\dot \beta_i}-m_i =\frac{p^2}{m_i}\left[ \frac{1}{2} - H \frac{\beta_i }{\dot \beta_i}\left( 1-\frac{\Gamma_{i}}{\Theta} \right) \right],
    \end{equation}
    and it is readily checked that  the solution $\alpha_i - m_i \beta_i = {constant}$, with the right hand side ({\it r.h.s} providing the solution:
    \begin{equation}\label{TNR}
         \frac{\dot {T_i}}{T_i}=-2 \frac{\dot a}{a} + \frac{2}{3} {\Gamma_{i}},
    \end{equation}
    which is the same macroscopic law as given by (\ref{temp2}). 
    
As an illustration, let us consider the phenomenological law, ${\Gamma_{(i)}} =3\beta_i H$, where $\beta_i=constant$. In this case, by choosing the present day scale factor, $a_{0i}=1$, it is immediate to obtain from (\ref{TNR}):
\begin{equation}\label{TNR1}
T_{i}=T_{0{i}}a^{-2(1 - \beta_i)} \Leftrightarrow T_i=T_{0i} (1+z)^{2(1 - \beta_i)},
\end{equation}
where in the second equality the redshift parameter was defined by $z\equiv a^{-1} - 1$. For  $\beta_i=0$ (no particle production), the usual equilibrium temperature law for a non-relativistic decoupled component is recovered.

As one may check, in terms of the redshift, the general solution $T_i (z)$ for a nonrelativistic fluid endowed with an arbitrary ``adiabatic'' particle creation rate  reads: 
\begin{equation}\label{TNR2}
T_i=T_{0i}(1+z)^{2}e^{\frac{1}{3}}{\int^{z}_0{\Gamma_{(i)}(z')\frac{dt}{dz'}dz'}}.
\end{equation}

{\item  The relativistic domain ($m_i << T_i$)}

In this case equation (\ref{TKG}) the above equation becomes:
\begin{equation}
 \frac{\dot \alpha_i}{\dot \beta_i}=E_i \left[1-\left( 1-\frac{\Gamma_{(i)}}{\Theta}\right)\frac{\dot a}{a}\frac{\beta_i }{\dot \beta_i} \right],
 \end{equation}
which leads to the solution $\dot \alpha_i =0$ (null chemical potential) while $\beta_i = 1/T_i$ thereby recovering the non-equilibrium thermodynamic result [see Eq.(29)] 
\begin{equation}\label{TR1}
\frac{\dot T_i}{T_i}=-\frac{\dot a}{a}+\frac{\Gamma_{i}}{3}.
\end{equation}
Again, for $\Gamma_{i} = 3\beta_i H$, the solution of the above equation reads:
\begin{equation}\label{TR2a}
T_i=T_{0i}a^{-(1 - \beta_i)} \Leftrightarrow T_i=T_{0i}(1+z)^{(1 - \beta_i)},
\end{equation}
a result to be compared with the non-relativistic solution (\ref{TNR2}). Different from the equilibrium case ($\nu_i=0$) this is a non-linear law. As physically expected, for a  given value of $z \neq 0$, the temperature is smaller than in the standard $\Lambda$CDM model. In the case of  CMB, the current value of the temperature has been fixed with great precison by the FIRAS-COBE and recalibrated by the WMAP \cite{Mather99,Fixen09}. It is also worth noticing that ($\ref{TNR1}$) has been extensively used in CMB studies related to Sunyaev-Zeldovich \cite{ZS69,SZ70} and excitation states of interestellar molecules like C, CN and CNO \cite{Lima97,TempL,LoSecco2001,Cui2005,Mol2002,Luzzi2009,Norte2011,Jetzer2011,Muller2013,Hurier2014,KK2015}.   By fixing the constant at the present time, the general solution of the above equation can be written as:
    \begin{equation}\label{TR2}
        T_i=T_{0i} \left( \frac{a_0}{a} \right) e^{{\frac{1}{3}\int^{t_o}_{t}{\Gamma_{(i)}(t')}{dt'}}}.
    \end{equation}
    In the simplest but interesting case, the creation rate $\Gamma_{(i)}$ remains constant for a given cosmic time interval. This kind of situation may happens at the early inflation phase or at late times of the evolution.  By defining $\Delta t= t_f - t_i$ one finds the general solution:
    \begin{equation}\label{TR3}
        T_f=T_i\left(\frac{a_i}{a_f}\right)e^{\frac{\Gamma_{(i)}}{3}(t_f - t_i)}.
    \end{equation}
    
    Now, in terms of the cosmic redshift, the general solution of the temperature  law (\ref{TR2a})  for the CMB fluid endowed with ``adiabatic'' photon creation is given by %(see also discussions below () and (\ref{temp2})):
    \begin{equation}\label{NEQT}
        T=T_0(1+z)e^{{\frac{1}{3}}{\int^{z}_0{\Gamma_{(i)}(z')\frac{dt}{dz'}dz'}}},
    \end{equation}
    and, as should be expected, for $\Gamma_{(i)}=0$, the same equilibrium result is recovered.
    It is also useful to show how the above temperature law (\ref{TR1}) for massless particles is compatible with the radiation thermal equilibrium relations coming out from the  kinetic approach. By eliminating $\Gamma_{(i)}$ from the balance equations (\ref{KNB}) and (\ref{ECLK})  it follows that:
    \begin{equation}
    \frac{\dot \rho_i}{\rho_i + p_i} = \frac{\dot n_i}{n_i} \equiv \Gamma_i - \Theta
    \end{equation}
    and since $n_i\propto T^{3}_{(i)}$, we see that for $p_i=\rho_i/3$, then, $\rho_i \propto n^{4/3}_{i}$, and also 
    \begin{equation}
         \frac{\dot \rho_i}{\rho_i}= 4\frac{\dot T_i}{T_i} \,  \Rightarrow \, \rho_i \propto T^{4}_{(i)}.
    \end{equation}
    The above equilibrium relations were first determined based on irreversible thermodynamics, but now it has been recovered from a kinetic approach. It means that under``adiabatic" conditions particles are created but the energy density and concentration as a function of the temperature are given by the same expressions obeyed by the states of  equilibrium, only the time dependence of each one are different in comparison with the equilibrium evolution. Indeed, by using this result one may show that the spectrum of radiation is also preserved in the course of the cosmic evolution (see \cite{Lima97} for a preliminary deduction). A more rigorous deduction for a decoupled mixture with creation of massless quantum particles (bosons and fermions) and the associated spectrum  will be discussed below.
    
\end{itemize}

%\begin{figure*}\label{fig2}
%\centerline {\psfig{figure=Fig2.eps,width=3.9truein,height=2.8truein}}
%\vspace{-1.4cm}\begin{minipage}[t]{6in} \caption{Density contrast for a CCDM cosmology emulating the $\Lambda$CDM model,   as discussed in \cite{LSC2016}. 
%$\delta^{eff}_{(m)}=\delta\rho^{eff}_{(m)}/{\rho^{eff}_{(m)}}$ for different values of the free parameter ($\sum_i{\alpha_{i}}$). 
%Blue line  is obtained for the best fit value from SNe Ia data. It reproduces exactly the  standard $\Lambda$CDM prediction for the nonrelativistic density contrast. We also stress  that  $\delta_{(meff)}$ describes only that portion of the  created components (baryons + CDM), which are able to appear as a clustered matter. }
%In this effective $\Lambda$CDM cosmology, the rigid vacuum medium corresponds to the smooth part, that is, $\rho_{T}=\rho^{meff} + \rho^{eff}_{(v)}$. Therefore, the CMB secondary anisotropies, like the Integrated Sachs-Wolfe (ISW) effect, may provide a crucial test beteween CCDM cosmology and $\Lambda$CDM. The main reason for that is the\, predicted existence of ``adiabatic" creation of photons, which is not considered in the standard $\Lambda$CDM cosmology (see text).}
%It remains gravitationally inert (unclustered) due to the creation pressure.}
%$P _{c}=-\rho^{eff}_{smooth} = -\sum_i{\alpha_{i}}\,\rho_{c0}$.}
%\end{minipage}
%\end{figure*}

\section{Cosmology with Creation of Baryons, CDM, CMB Photons and Neutrinos}

As remarked before (see introduction), a new scenario emulating  the $\Lambda$CDM model, the so-called CCDM cosmology is based  on the creation of cold dark matter alone \cite{LJO2010}. Some attempts to consider creation of some dominant components (baryons + CDM) with different creation rates, were  also discussed in the literature \cite{LSS09,LSC2016}. Now we show that the extended CCDM model macroscopically proposed in \cite{LSC2016} which has an evolution  equivalent to $\Lambda$CDM  both at the background (cosmic history) and perturbative levels (linear and nonlinear), can also be formulated in a natural way based on the kinetic theoretical formulation as developed in the previous section (see also Appendix A). 

\subsection{The Extended CCDM Model}

For each decoupled component, the general ratio $\Gamma_i/\Theta = \alpha_i\rho_{co}/\rho_i$, now takes the following form:
\begin{align}
\label{Gamma} 
\frac{\Gamma_b}{3H}=\alpha_b \frac{\rho_{co}}{\rho_{b}}, \,\,\,\,\,\frac{\Gamma_{dm}}{3H}=\alpha_{dm} \frac{\rho_{co}}{\rho_{dm}}, \\ \frac{\Gamma_{r}}{3H}=\alpha_{r} \frac{\rho_{co}}{\rho_{r}}, \,\,\,\,\,\, \frac{\Gamma_{\nu}}{3H}=\alpha_{\nu} \frac{\rho_{co}}{\rho_{\nu}},
\end{align}
where $\alpha_b$,  $\alpha_{dm}$, $\alpha_{r}$ and  $\alpha_{\nu}$ are constant parameters, while $\rho_{co}$ is the present day value of the critical density. Similarly, for each component, the creation pressure, $P_{ci}= - (\rho_i + p_i)\Gamma_i/3H$, reduces to: 
\begin{align}
P_{cb}= -\rho_b\Gamma_b/3H=-\alpha_b\rho_{co},\\ 
P_{dm}=-\rho_{dm}\Gamma_{dm}/3H=  -\alpha_{dm}\rho_{co},\\
P_{cr}={-\frac{4}{3}}\rho_{r}\Gamma_{r}/3H=  -\frac{4}{3}\alpha_{r}\rho_{co},\\
P_{c\nu}={-\frac{4}{3}}\rho_{\nu}\Gamma_{\nu}/3H=  -\frac{4}{3}\alpha_{\nu}\rho_{co},
\end{align}
where for simplicity we have also assumed massless neutrinos. Note also that each creation pressure is modulated by its specific creation parameter $\Gamma_i$,  as defined by equations (66)-(67). 

At late times all these components are decoupled and radiation and neutrinos are subdominant in the deep matter phase. Although dynamically irrelevant at zero order, it is well known that CMB photons (and neutrinos) play  an important role in the perturbative approach both from a thermodynamic and kinetic viewpoints.  For a while we neglect such components. In this case, $P_{cT}= - (\alpha_b + \alpha_{dm})\rho_{co}$, so that the total creation pressure  depends only on the effective creation rate parameter, $\alpha_{eff}= \alpha_{dm} + \alpha_b$.

Now, by combining Friedmann equation
\begin{equation}
8\pi G(\rho_{dm} + \rho_b) = 3H^{2},
\end{equation} 
with the energy conservation law for both components one finds:
\begin{equation}
\label{HzFlat}
 \left(\frac{H}{H_0}\right)^2= \Omega_{meff}(1+z)^3 + \alpha_{eff},
\end{equation}
where $\Omega_{meff}= \Omega_{dm} + \Omega_{b} - \alpha_{eff} \equiv 1 - \alpha_{eff}$ is the clustered matter. Note  that $\alpha_{eff}$ allows a reduction of the dark sector, thereby emulating the $\Lambda$CDM dynamics with  $\alpha_{eff}=\alpha_b + \alpha_{dm}$. Actually, by integrating  (\ref{HzFlat}) we obtain:

\begin{equation}\label{FS1}
a(t)=\left(\frac{1-\alpha_{eff}}{\alpha_{eff}}\right)^{1/3}
\sinh^{\frac{2}{3}}\left(\frac{3H_{0}\sqrt{\alpha_{eff}}
}{2}t\right),
\end{equation}
which  is identical to that predicted by the standard flat $\Lambda$CDM model with only one new free dynamic parameter, $\alpha_{eff}$. The analogy is perfect by identifying $\Omega_{\Lambda} \equiv \alpha_{eff}$.  It is widely known that $a(t)$ is not directly observable. However, for $a(t) = a(t_0 )=1$,  the age of the Universe today, $t_0$, can be calculated. In this way, a lower bound on $\alpha_{eff}$ can be obtained when we compare it with the values of the oldest objects in our galaxy or even at high redshifts. Naturally, this can also be done using expression (\ref{HzFlat}) for the Hubble parameter as  discussed long ago for the $\Lambda$CDM model \cite{Lima99}.  

Of course, when the creation of photons and neutrinos are not taken into account, such a reduced dark sector scenario mimics the cosmic concordance model  from a dynamic viewpoint, and its thermodynamic behaviour is not modified. However, when the creation of CMB photons and neutrinos are added, the value of H(z) as given by (\ref{HzFlat}) is modified. In particular,  at the level of the EFE, the new effective creation parameter $\bar \alpha_{eff}$ does not appear additively so that  $\bar \alpha \neq \alpha_b + \alpha_{dm} +\alpha_{r} + \alpha_{\nu}$, and, as such, the model dynamics also require more than one free parameter. Of course, at the matter dominated phase the CMB temperature law depends only of $\alpha_r$, as should be expected from (\ref{temp3}) [see also the kinetic derivation (\ref{TR1})]. In this case,  the equilibrium redshift and other relevant properties of the photon-baryon fluid are slightly modified.

An interesting effect to the large scale structure and CMB anisotropies (see subsection {\bf VB})  is related with the $\alpha_{eff}=\alpha_{dm} + \alpha_b$ driving the evolution of the non-relativistic matter density perturbation. In this case, the growing mode solution in the matter dominated phase in terms of the scale factor can be expressed as \cite{LSC2016}:

\begin{equation}\label{deltam}
\delta_{meff}(a) = C({\bf x})aF\left(\frac13, 1; \frac{11}{6};
-\frac{\alpha_{eff}a^3}{1-\alpha_{eff}}\right), 
\end{equation}
where  $C({\bf x})$ is an integration constant $F={}_2F_1(a, b, c,z)$  is the Gaussian hypergeometric function. As should be expected, if the net creation parameter $\alpha_{eff}=\alpha_{dm}+\alpha_b = \sum_i\alpha_i\rightarrow 0$ so that the results of the standard Einstein-de Sitter model are recovered $({\delta^{eff}_{(m)} \propto a},\, \Omega^{eff}_{(m)} \rightarrow 1)$. 

\begin{figure}
\includegraphics[width=3.5truein,height=3.0truein]{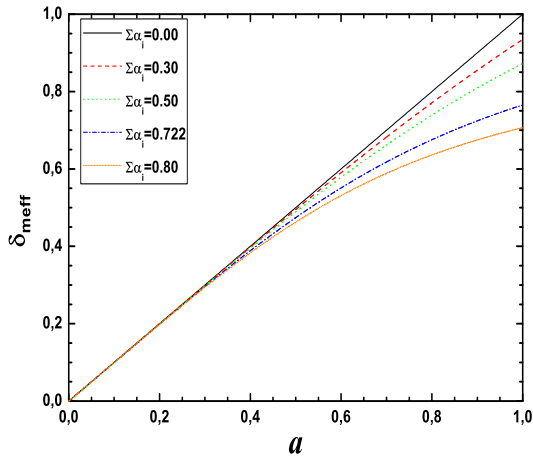}\caption{Evolution of the effective matter density contrast in the extended CCDM model (for different values of $\alpha_{eff}$ as a function of the scale factor. 
The blue line is obtained for the best fit value from SNe Ia data to the unique effective free parameter, $\alpha_{eff}=\alpha_{dm} + \alpha_b$. It reproduces exactly the  standard $\Lambda$CDM prediction for the nonrelativistic density contrast and also to the transition redshift. Note that $\delta_{(meff)}$ describes only that portion of the created nonrelativistic components (baryons + CDM), which is able to appear as clustered matter. This extended CCDM model is  different from \cite{LJO2010, Waga2014a} since it also includes the created subdominant components (CMB + neutrinos) thereby slightly changing the temperature law and others relevant properties of the photon-baryon fluid.}
\label{DC}
\end{figure}
In figure 1, we display the evolution of the contrast density as a function of the dynamically relevant created components \cite{LSC2016}.  It is interesting that the best fit of the effective creation parameter, $\alpha_{eff}=\alpha_{dm} + \alpha_b$, from SNe Ia data is guaranteeing two nontrivial results, namely: the same evolution for the density contrast (blue line) and also the same value of the transition redshift as predicted by the flat $\Lambda$CDM model.  In addition, the modified CMB temperature evolution from the same current value as given by the COBE data and recalibrated by WMAP, $T_0=2.72548 \pm 0.00057$K, points to a new physics close to $\Lambda$CDM model, potentially, modifying the relatively smaller value of $H_0$ as predicted by CMB. In other words,  creation of photons satisfying the modified temperature law  suggests a crucial test in the thermal sector involving distortions, CMB temperature anisotropies and the value of $H_0$ itself, even considering that the same $\Lambda$CDM dynamics is preserved. At this point, it is natural to ask how distortions and  CMB anisotropies would be investigated in this enlarged context emulating the $\Lambda$CDM dynamics but not its thermodynamics and kinetic approach. 

\subsection{Extended CCDM Cosmology and CMB Distortions: A Case for SZE}

To begin with we investigate whether the zero-order spectrum is preserved. Although CMB distortions have already been partially investigated in a recent separated paper \cite{LTS2021}, this is needed because the underlying connections with the present framework were not properly discussed. Photons and neutrinos (massive or massless) have currently different temperatures. Now, the interest for distortions in this framework will be illustrated with the Sunyaev-Zeldovich effect (SZE). In the next subsection, its  cross-correlation with CMB anisotropies will be discussed as a possible crucial test probe for CCDM and $\Lambda$CDM cosmologies.  

Let us now consider an arbitrary decoupled massless component (bosonic or fermionic) at temperature $T_{i}$, $i=r,\nu$. The general relativistic equilibrium distribution function for a massless dilute quantum gas with zero chemical potential takes the form: 
\begin{equation}
    \label{D1}
    f^{0}_{(i)}(t,p) = \frac{1} {e^{\frac{p}{T_i}} - \xi_i}
    %f^{0}(t,p) = \frac{1} {e^{\left(\frac{p}{T}\right)} - \xi}
\end{equation}
where $p$ is the physical momentum and for bosons ($\xi_i = +1$), while for fermions ($\xi_i = -1$). In this section we are  following closely the notation  of the textbook \cite{DOD2003} for CMB photons. Our basic aim here is to demonstrate that both quantum equilibrium distributions above are preserved in the course of the expansion when the creation of massless particles happens under ``adiabatic" conditions. Of course, $f^{0}_{(i)}$ is also the solution without creation since the collisional term is also identically zero for equilibrium states.  

Now, by assuming  ``adiabatic'' creation with a rate $\Gamma_i$, the modified collisionless equation (\ref{MSBE}) for each component can be rewritten as:

\begin{equation}
    \label{D2}
\frac{\partial f^{0}_{(i)}}{\partial t} + \left(-H + \frac{\Gamma_i}{3} \right) p \frac{\partial f^{0}_{(i)}}{\partial p} = 0.
\end{equation}             
The first term in the equation above can be rewritten as:  
\begin{equation}
    \label{D3}
    \frac{\partial f^{0}_{(i)}}{\partial t} = \frac{\partial f^{0}_{(i)}}{\partial T_i} \dot{T_i},
\end{equation}
while the equilibrium distribution form (\ref{D1}) implies that:
\begin{equation}\label{D4}
 \frac{\partial f^{0}_{(i)}} {\partial T_i} = - \frac {p}{T_i} \frac{\partial f^{0}_{(i)}}{\partial p} \Rightarrow \frac{\partial f^{0}_{(i)}}{\partial t} = -\frac{\dot T_i}{T_i} p \frac{\partial f^{0}_{(i)}}{\partial p},
\end{equation}
where in the second equality above the result in (\ref{D3}) has been used.
%\begin{equation}
%\frac{\partial f_i}{\partial t} = \frac{p_{(i)}}{T} \frac{\partial f_i}{\partial p_{(i)}} %\frac{dT}{dt} 
%\end{equation}
Therefore, by inserting the above derivative into the MBE (\ref{D2}) we obtain:
\begin{equation}
    \left[-\frac{T_i}{T_i} - \frac{\dot a}{a} + \frac{\Gamma_i}{3} \right] p \frac{\partial f^{0}_{(i)}}{\partial p} = 0,
\end{equation}
thereby providing the temperature law for quantum massless particles: 
\begin{equation}\label{TCMBF}
\frac{\dot T_i}{T_i} = -\frac{\dot a}{a} + \frac{\Gamma_i}{3}\,\, \Leftrightarrow\,\, T_i(t)a(t)N_i(t)^{-\frac{1}{3}} = constant. 
\end{equation}
Note that the above equation is the same temperature law deduced before for a dilute ultra-relativistic gas ($T_i >> m_i$) of non-quantum point particles [see Eq. (\ref{TR1})]. Note also that in the second equality above $\Gamma_i = {\dot N_i}/{N_i}$ has been  used. It thus follows that the equilibrium spectrum (\ref{D1}) for massless quantum particles is preserved in the course of the expansion regardless of the value of $\xi_i =\pm 1$. The price to pay is that the temperature law for each component (CMB photons or neutrinos) is modified by creation rate of the massless decoupled component. Thus, an interesting question here is how such a preserved  blackbody spectrum  preserved  will be  slightly distorted in the course of the expansion? Let us discuss that with a simple example.

  Spectral distortions may be provoked by collisional processes like Compton scattering (C), double Compton (DC), and Bremsstralung (BR). Such processes usually involve a redistribution of photons over frequencies and sometimes readjustments on the photon number \cite{Ota2019,AC2020,ML2021,Y2021}.   In order to exemplify that let us discuss the CMB distortions provoked by the inverse Compton scattering  at low redshifts. In cosmology there are two important processes. The first is widely known as the (thermal) Sunyaev-Zeldovich effect (SZE) after their seminal papers \cite{ZS69,SZ70}. This means that the modified collisionless Boltzmann equation with creation (\ref{MSBE}) must be used in its complete form, that is, by including the required sources of spectral distortion. In the case of SZE, for instance, such a study must start based on the extended equation including the Compton scattering: 

\begin{equation}\label{SD1}
 \frac{df}{dt}\equiv \frac{\partial f}{\partial t} - H p \frac{\partial f}{\partial p}  + \frac{\Gamma}{3} p\frac{\partial f}{\partial p} = \frac{df}{dt}|_C,   
\end{equation}
where the new contribution in the right-hand-side is the corresponding collisional term. 
 
The  thermal SZE is the spectral distortion of CMB caused by inverse Compton scattering of photons when transverse the hot electrons  of an ionized gas ($T_e \sim 10^{7-8}K$) across the line of sight. The total number of photons in such elementary processes is conserved and its effect is quite simple, namely: CMB photons are up-scattered by the hot electrons thereby depopulating the Rayleigh-Jeans low frequency region of the spectrum. As a result, the scattered photons move to the high energy side of the photon distribution. By assuming an initial perfect blackbody CMB spectrum, the  net effect after scattering  is that the spectrum is slightly distorted. In the standard treatment (no photon creation) the SZE is simplified because it does not depend on the redshift. However,  such a condition  is violated in the present framework because the standard $\Lambda$CDM temperature law is not obeyed [see, for instance, the kinetic law  (\ref{TCMBF})]. Therefore, if properly studied based on the above modified Boltzmann equation with creation, the SZE may become a crucial test confronting $\Lambda$CDM and the present extended CCDM cosmology. 
 
 It should also be recalled that the motion of clusters relative to the Hubble flow also produce a "kinematic SZ effect" (KSZE). This effect is usually much smaller than the thermal SZE effect, but it is also quite relevant in the present context since the KSZE can be used to determine  the behavior of clusters and the Hubble constant itself. Naturally, a detailed investigation of CMB secondary distortions based on the SZE by taking into account ``adiabatic" photon creation process and inverse Compton scaterring as described in ($\ref{SD1}$) 
 is beyond the scope of this paper, and will be discussed in a forthcoming communication.

\subsection{Extended CCDM and CMB anisotropies: The case for ISW}

 As discussed in the previous sections, ``adiabatic" creation by the smooth universe cannot by described  as a source of CMB distortions because the Planckian spectrum is preserved. The only net effect of such a creation process to CMB is a modification of the temperature redshift relation. This means that the perturbation of the photon distribution function may be written in the standard way \cite{DOD2003}   
%\begin{equation}\label{DFF}
%f(p, {\hat n}, {\bf r},t) = \left[exp \left{\frac{p}{T(t)[1 + \Theta (t, {\hat n},{\vec r})]}\right}} - 1 \right]^{-1},  
%\end{equation}
\begin{equation*}
    f(t, {\bf r}, p, \hat{p}) = \left[\exp \left\{\frac{p}{T(t) [1 + \Theta (t, \vec{r}, \hat{p})} \right\} - 1 \right]^{-1}
\end{equation*}
where  $T(t)$ is the zero order-temperature with photon creation and $\Theta = \delta T/T\, (t,{\bf r},\hat{p})$ is the small fractional dimensionless temperature perturbation observed in the direction of the unit vector ${\hat p}$  on the sky at the time $t$ and position ${\bf r}$.  In addition, since the created photons share the same temperature of the already existing ones, all collision terms in the presence of ``adiabatic" creation must be proportional to $\Theta$ and other perturbatively small quantities. This happens because the induced creation contribution  by the expanding universe is not equivalent to a collisional term. Thus, at zero-order the Boltzmann equation equation is also isotropic as happens for any FLRW metric, as for instance, $\Lambda$CDM model. 
%for $\Gamma_r=3\alpha_r H$, 
%On the other hand, the complete hierarchy of the perturbed Boltzmann equations with the collisionless Boltzmann equation (36) is modified in a very simple way, namely:  the value of $H$ is replaced by an effective value, $H\,\, \rightarrow \,\, H_{eff}=  (1-\alpha_r)H$.  Note also that the zeroth Boltzmann equation is also fully equivalent  to
%\begin{equation}\label{Temp6}
%    \frac{\partial}{\partial \eta}\left[Ta^{1-\alpha_{r}}\right]=0, 
%\end{equation}
%which for $\alpha_r =0$ boils down to the standard $\Lambda$CDM result, informing us that temperature in the universe is inversely proportional to the scale factor. Thus $\alpha_r \neq 0$ means that the temperature law departs from the standard evolution and must affect the anisotropies.   

In a point of fact, some authors already discussed CMB temperature anisotropies in models with creation of CDM, but not in the general framework presented here (see section V), which is dynamically equivalent to $\Lambda$CDM. 
%Naturally, in comparison with the  $\Lambda$CDM model,  a cosmology  with $\Omega_{DE}\equiv 0$ endowed with creation of CDM alone also changes the anisotropy pattern (angular power spectrum) of CMB. However, although interesting and motivating, this kind of analysis seems to be inherently incomplete because the creation of the remaining components were not taken into account based on the full Boltzmann hierarchy. In particular, photon creation was not considered. 
In \cite{nunes2016}, for instance, it was assumed that  baryons, and photons are conserved as in the standard $\Lambda$CDM treatment and the influence of neutrinos was also not considered. Three different phenomenological expressions of the creation rate $\Gamma_{dm}$ were assumed. In their simplest model (MI), the creation rate was defined by $\Gamma_{dm} = 3\alpha_{dm}H$ (the authors used $\beta$ instead of $\alpha_{dm}$). The effects on  CMB TT  power spectrum (and also to CMB EE)  were obtained and compared with the theoretical predictions of the $\Lambda$CDM (see their Figs. 4 and 5). As should be expected, due to the excess of CDM in comparison to baryons, significant  deviations from $\Lambda$CDM were obtained for $\alpha_{dm} \geq 0.05$.  

Nevertheless, although physically interesting their results cannot be considered definitive by the following reasons (i) the unperturbed model (cosmic history), although presenting a transition from acceleration to a decelerating regime, does not reproduce the  $\Lambda$CDM dynamics, and (ii) the complete hierarchy of the perturbed Boltzmann equations for all components with creation were not considered. Note that the first condition is somewhat desirable because of the recognised successes of $\Lambda$CDM for many cosmic probes. Implicitly, it also means  that any  realistic cosmology must be as close as possible to $\Lambda$CDM model, but being slightly different in order to point out a new route to handle the tensions and also shed some light in the theoretical puzzles (coincidence and $\Lambda$ problems). 

In  this context, let us now highlight some new physical results predicted by the extended CCDM model and its comparison with $\Lambda$CDM.  To begin with,  we stress  that the scale factor $a(t)$, as given by (\ref{FS1}) has the same expression of the $\Lambda$CDM model. In addition, the evolution of the density contrast is also the same of $\Lambda$CDM [see Eq. (\ref{deltam}) and Fig. 1].  Both results plus the modification of the temperature, in principle,  are very significant to CMB anisotropies. To show that consider now the perturbed metric in the potential conformal Newtonian gauge. The temperature anisotropy provided by the change in the Newtonian potential along the line of sight since the LSS until the present day is often referred to as integrated Sachs-Wolfe (ISW) effect. 
%of the  $\Lambda$CDM model, but now without any kind of dark energy. 

Now, in order to understand easily the forward step given here, we first assume that CMB photons are not created. In this way, the ISW effect assumes the standard expression \cite{DOD2003,ME2004}  
\begin{equation}\label{ISW}
\Theta_{ISW} \equiv \frac{\delta T({\bf \hat p})}{T_{0}}\, =\, {2}\,\int_{\eta_{LSS}}^{\eta_0}
           d\eta\; \frac{\partial\Phi({\bf r},\eta)}{\partial \eta}\,,
\end{equation}
%where $\theta (\eta, {\hat p}, {\vec r})$ is a unit vector in the direction   
where $T_{0}$ is the present day temperature  and the limits of integration ranges from the recombination (LSS) to the present time, respectively, whereas  $\Phi({\bf r},\eta)$ is the time-varying gravitational potential along the photon path. Recalling that only low redshifts are important to the above integral, we have also ignored the  suppression factor caused by the Thomson scattering \cite{Ho2008}. 

It is widely known that for nonrelativistic matter in the Einstein-de Sitter cosmology, $a(t)\propto t^{2/3}$,  the Newtonian gravitational potential is time-independent and the ISW is identically null. In contrast, the late time dominance of the vacuum energy density in the $\Lambda$CDM cosmology gives rise to a time-varying potential, and, as such, the ISW effect is different from zero \cite{ME2004,KS85} and have also been observed by different groups \cite{H2013,G2008,S2019}.  Hence, since the extended CCDM model driven by non-relativistic matter density plus its creation pressure is fully equivalent to $\Lambda$CDM, this means that the ISW effect is exactly the same of the standard cosmology whether photons are not created as assumed in \cite{nunes2016}. 

On the other hand, some reported observational results for the ISW are in contradiction with the $\Lambda$CDM prediction. For example, an excess signal of ISW effect has been reported by several authors based on cross-correlation between WMAP and catologues of quasars, clusters, supervoids and other surveys \cite{S2019,G2008,H2013} and confronted with the $\Lambda$CDM results. Recently, new constraints were derived cross-correlating  Planck's temperature maps with AGN and radio sources catalogues thereby obtaining a very positive detection of the ISW signal at 5$\sigma$ of significance level. In particular, this means that in the near future with more data and an improved  treatment of systematics for different surveys, potentially,  may provide an additional difficulty to the standard $\Lambda$CDM cosmology.

{In the same vein,  we recall that some recent studies are based on cross-correlations of CMB, Gamma-Ray background and the SZE effect. In our view, given the above results and regardless of the present status of $\Lambda$CDM concerning the quoted analyses (see Sulton\cite{S2019} for a short and nice review in the observational front),  it seems interesting to propose a crucial test involving the extended CCDM model and $\Lambda$CDM cosmology. The reason is very simple. As we have seen, the extended CCDM cosmology  has the same dynamics, but its thermodynamics is slightly different from $\Lambda$CDM. In particular, the SZE effect is not independent of the redshift as occurs in $\Lambda$CDM model (see discussion on the previous subsection). Moreover, the Boltzmann equation with photon production in CCDM model means that the first order perturbed Boltzmann equation for photons is also  slightly modified  in comparison with the standard $\Lambda$CDM treatment. On the other hand, since the analysis of the SZE is also modified by\,``adiabatic" photon creation, the cross-correlation between CMB and SZE, is the interesting statistical tool for a sharp test confronting CCDM and $\Lambda$CDM.}
%Of course, a more complete treatment along these lines are beyond the scope of the present paper. 

\section{conclusion}

In this paper we have investigated the {\it thermodynamic} and {\it kinetic} properties
of an arbitrary decoupled multi-fluid mixture endowed with gravitationally induced particle production of all components, in principle,  with different creation rates $\Gamma_{i}$.  The main results derived
here may be summarised in the following statements:

1) For each component, the efficiency of the phenomenon depends on the ratio $\Gamma_{i}/H$. Of course, for a given component, the process is negligible whether $\Gamma_{i} << H$.  The irreversible macroscopic results are valid for any FLRW geometry and also for values of $\Gamma_{iN}$ and  $\Gamma_{iS} \geq 0$. However, the kinetic counterpart was deduced only for the flat case ($k=0$) and adiabatic creation defined by $\Gamma_{iN}=\Gamma_{iS}$.

2) The whole process is irreversible but the gain of entropy in the ``adiabatic" case (the most interesting one from a physical point of view),  depends only on the  created particles ($S_i=k_B N_i$). This happens because $\Gamma_{iS} = \Gamma_{iN}=\Gamma_{i}$ so that $\dot \sigma_i = 0$ [see discussion in section III right below Eq. (\ref{dotsigma})]. For each decoupled component, this means that both the total entropy and the number of particles increase but the specific entropy (per particle), $\sigma_i = S_i/N_i$, remains constant.  

%iii) All kind of particles Physically, one may think that new particles of all components (with different creation rates)  spring-up into the spacetime because energy is being continuously supplied by the classical time varying gravitational field, a process of quantum origin which has been here kinetically described. 
%works like a ‘pump’ providing energy with the energy of the  supplied  energy to the quantum fields in such a way that particles  
%For each decoupled component, it was showed that the macroscopic process named ``adiabatic'' particle production has a consistent kinetic counterpart. In this kind of process the entropy increases but the specific entropy remains constant \cite{CLW92,LG92}.
3) The multi-fluid approach developed here is in fact a quasi-zero-order description, in the sense that the relativistic distributions has the same form of equilibrium. In particular, the CMB blackbody spectrum with creation is not destroyed in the course of the expansion. Therefore, at zero order, the extended  CCDM cosmology with creation of CDM, baryons, photons and neutrinos (see section Va)  is now dynamically described by $H(t)$ and the different creation rates $\Gamma_{i}$. These quantities $\Gamma_i$  affect the temperature law of each component.  The extra bonus of the extended CCDM cosmology is that dark  energy is not required ($\Omega_{DE}=0$) anymore thereby solving naturally the coincidence and $\Lambda$-problem. Particularly, the transition from a decelerating to an accelerating regime in the matter dominated phase is provided by the negative creation pressure of the baryonic and CDM components  [see Eqs. (\ref{EE2b}) and (\ref{PCiK})]. 

4) All the macroscopic results obtained in the irreversible macroscopic approach for the decoupled multi-fluid were recovered by the associated  kinetic treatment. 

%The temperature law was determined for two important limits, namely, the ultrarelativistic ($m_i << T_i$) and nonrelativistic ($m_i >> T_i$) domains.  The energy density of massless or ultrarelativistic particles follow the standard thermodynamic laws ($\rho_i \propto T_i^{4}, n_i \propto T_i^{3}$) as shown in the macroscopic case. In particular, this means that early conjectures showing that the CMB spectrum is not necessarily destroyed when the cosmic creation process happens under ``adiabatic'' conditions \cite{Lima97} has been kinetically confirmed, based on a more rigorous approach involving a multi-fluid mixture. This is an interesting aspect since all kind of particles should potentially be created by the expanding Universe in order to satisfy the equivalence principle. 

5) When photon creation is neglected it was shown that the ISW effect of the extended CCDM model is the same of $\Lambda$CDM cosmology. However, this result is modified when  CMB photons are created because (i) the temperature is modified, and (ii) the perturbed Boltzmann equation for CMB photons acquire an additional term.  In particular, this means that the standard Sunyaev-Zeldovich effect is not independent of the redshift as happens in the $\Lambda$CDM model.

%As it will be discussed elesewhere, a consistent approach to the SZE effect will require a proper modification of the standard Kompaneets equation to include photon creation  

6) The present analysis also open a new window to investigate the $H_0$ and $S_8$ tensions in virtue of twofold reasons: (i) The unperturbed model CCDM model has the same  $\Lambda$CDM dynamics (linear and nonlinear leves) powered by non-relativistic matter, (ii) The creation  of the remaining components (CMB photons and neutrinos)  changes slightly  the expansion history at early and late times (see discussions in section VB). Its physical consequences at the level of the $H_0$ and S8 tensions will be discussed with more detail in a subsequent paper. 

Finally, we also emphasise an interesting  aspect related to the spectral distortions and CMB anisotropies in the presence of ``adiabatic" photon creation. As discussed in section VB and VC,   
%At present, many authors believe that possible interactions of dark matter (DM) with standard model particles or even with dark energy can also be tested by exploring the CMB physics of spectral distortions. As extensively discussed here, by adopting a reasonable ansatz to the creation rate $\Gamma_{i}$ 
the predictions of the extended CCDM cosmology with $\Omega_{DE}=0$ must not only be compared with the observations but also confronted with the ones of the $\lambda$CDM model. In principle, the rationale and soundness of gravitationally induced particle production requires much more work and analysis based on the upcoming data, in particular, for prospecting the main consequences for the angular power spectrum and CMB distortions, as well as their cross correlations with different surveys (subsections Vb and Vc). As argued there, since the analysis of the SZE is also modified by\,``adiabatic" photon creation, the cross-correlation between CMB temperature maps and SZE (and other surveys) seems to be the interesting statistical tool for a crucial and accurate test confronting the extended CCDM and $\Lambda$CDM models. 
\vspace{0.3cm}
%It should be stressed that the kinetic treatment  has been discussed only to the case of a flat geometry. The unified general case ($k=0, \pm 1$), including a multicomponent fluid,  will be discussed elsewhere.

\begin{acknowledgments}  
\noindent JASL is partially supported by the National Council for Scientific and Technological Development (CNPq) under grant 310038/2019-7 and , CAPES (88881.068485/2014), and
FAPESP (LLAMA Project No. 11/51676-9). SRGT also acknowledges the support of CNPq.
\end{acknowledgments}

\appendix
\section {Boltzmann Equation and ``Adiabatic" Creation}

Let us discuss with more detail how the standard collisionless relativistic Boltzmann equation is modified in the presence of ``adiabatic" matter creation. The main aim here is to derive the ``mass shell" Boltzmann equation (\ref{MSBE}) by starting from the  covariant form (\ref{Beq1}): 
\begin{equation}\label{A1}
{\mathcal L}(f_i)\equiv P^\mu_{(i)}\frac{\partial{f_{(i)}}}{\partial{x^\mu_{(i)}}}-\Gamma^{\mu}_{\alpha \beta}P^\alpha_{(i)}P^\beta_{(i)}\frac{\partial f_{(i)}}{\partial P^\mu_{(i)}} + {\mathcal P_{Gi}}(x^{\mu}_{(i)},P^{\mu}_{(i)}) = 0,
\end{equation}
 The undefined  quantity, ${\mathcal P}_{Gi}$, is assumed proportional to both terms $\Gamma_{i}/{\Theta}$ and $\Gamma^{\mu}_{\alpha \beta}P^\alpha_{(i)}P^\beta_{(i)}\frac{\partial f_{(i)}}{\partial P^\mu_{(i)}}$  [see discussion below (\ref{Beq1})]. Like the expansion itself (second term), the form adopted above for  ${\mathcal P}_{Gi}$ reflects the fact that ``adiabatic" matter creation is also a purely gravitational effect. Thus, (\ref{A1}) takes the form:

\begin{equation}\label{A2}
P^\mu_{(i)}\frac{\partial{f_{(i)}}}{\partial{x^\mu_{(i)}}}-\left(1- B\frac{\Gamma_{i}}{\Theta}\right)\Gamma^{\mu}_{\alpha \beta} P^\alpha_{(i)}P^\beta_{(i)}\frac{\partial f_{(i)}}{\partial P^i\mu_{(i)}}  = 0,
\end{equation}
%+ {\mathcal P_{Gi}}(x^{\mu}_{(i)},P^{\mu}_{(i)})
where $B>0$ is a pure number of the order of unity. It must be determined in such a way that all the ``adiabatic" balance equations with creation are kinetically reproduced. Note also that  the ``mass shell'' constraint, $g_{\mu\nu}{P^\mu}_{(i)} {P^\nu}_{(i)}= m_{(i)}^2$ implies that  $f_{(i)}\equiv f_{(i)} (t,P^i_{(i)})$  with the above equation reducing to: 
\begin{equation}\label{A3}
P^{0}_{(i)}\frac{\partial{f_{(i)}}}{\partial{t}}-2HP^{0}_{(i)}\left(1- B\frac{\Gamma_{i}}{\Theta}\right)P^i_{(i)}\frac{\partial f_{(i)}}{\partial P^i_{(i)}}  = 0,
\end{equation}
where we have replaced the values $\Gamma^i_{0j}=\Gamma^i_{j0}= H\delta^{i}_j$ from ($\ref{Csymb}$). In addition, from spatial homogeneity and isotropy condition and mass shell condition, the distribution function of each decoupled component is a function of the  time and energy $P^{0}_{(i)} = E_i$ (or, equivalently, the modulus of the momentum of the i-th component $P=P_{(i)}$). Thus, we may rewrite the above expression as:

\begin{equation} \label{A4}
\frac{1}{E_i}{\mathcal L}(f_i)\equiv\frac{\partial{f_{(i)}}}{\partial{t}}-2H\left(1- B\frac{\Gamma_{i}}{\Theta}\right)P_{(i)}\frac{\partial f_{(i)}}{\partial P_{(i)}} = 0.
\end{equation}
Following standard lines, let us rewrite the above equation in terms of the physical momentum, ${p}_{(i)} \equiv a(t) P_{(i)}^i$: In this case, the time derivatives of the distribution function $f_i(t,P_{(i)})$ and $f_i(t,p_{(i)})$ are related by: 

\begin{equation}
	\frac{\partial f_{(i)}}{\partial t} (t, P_{(i)}) = \frac{\partial f_{(i)}}{\partial t} (t, {p}_{(i)}) + H{p}_{(i)}\frac{\partial f_{(i)}}{\partial {p}_{(i)}}.
\end{equation}
Now, inserting such results into (\ref{A4}) it follows that:
\begin{equation}
 \frac{1}{E_i}{\mathcal L}[f_{(i)}]\equiv\frac{\partial{f_{(i)}}}{\partial{t}}-H\left(1- 2B\frac{\Gamma_{i}}{\Theta}\right)p_{(i)}\frac{\partial f_{(i)}}{\partial p_{(i)}}  = 0.   
\end{equation}
Finally, by comparing with the Liouville operator (\ref{MSBE}) from which the balance equations for ``adiabatic" creation were kinetically calculated thereby reproducing all the macroscopic results [cf. Eqs. (\ref{PCL2}) and (\ref{DivS1}) in section III with the corresponding Eqs. (\ref{KNB}) and (\ref{KSB}) in section IV], we may  conclude that the only possible value of the undetermined pure number in (\ref{A2}) is $B=1/2$.

\end{document}